\documentclass[
preprint,
]{jasatex}

\usepackage{graphicx}
\usepackage{dcolumn}
\usepackage{amsmath,amsfonts}
\usepackage{bm}
\usepackage{amssymb}
\usepackage[english]{babel}
\usepackage{epsf,exscale,times}
\usepackage{varioref}
\usepackage{ams}
\usepackage{graphicx}
\usepackage[bf]{subfigure}
\usepackage{epsfig}
\usepackage{wrapfig}
\usepackage{lettrine}
\usepackage{setspace}
\usepackage{fancyhdr}
\def\thesection{\arabic{section}}

\def\@sections#1{\@startsection {section}{1}{\z@}                   
                {\large\centering}{\nouppercase{#1}}}

\begin{document}
%

\title[Compressive sensing beamforming]{Compressive sensing based beamforming for noisy measurements}

\author{Siyang Zhong}

\affiliation{%
Department of Mechanics and Aerospace, College of Engineering, Peking University, Beijing, 100871, China.}
\vspace{15mm}

\author{Xun \surname{Huang}}
 \email{huangxun@pku.edu.cn}
 \affiliation{%
huangxun@pku.edu.cn, State Key Laboratory of Turbulence and Complex Systems, College of Engineering, Peking University, Beijing, 100871, China}

\date{\today}

\begin{abstract}
Compressive sensing is the newly emerging method in information technology that could impact array beamforming and the associated engineering applications. However, practical measurements are inevitably polluted by noise from external interference and internal acquisition process. Then, compressive sensing based beamforming was studied in this work for those noisy measurements with a signal-to-noise ratio. In this article, we firstly introduced the fundamentals of compressive sensing theory.  After that, we implemented two algorithms (CSB-I and CSB-II). Both algorithms are proposed for those presumably spatially sparse and incoherent signals.
The two algorithms were examined using a simple simulation case and a practical aeroacoustic test case. The simulation case clearly shows that  the CSB-I algorithm is quite sensitive to the sensing noise. The CSB-II algorithm, on the other hand, is more robust to noisy measurements. The results by CSB-II at $\mathrm{SNR}=-10\,$dB are still reasonable with good resolution and sidelobe rejection. Therefore, compressive sensing beamforming can be considered as a promising array signal beamforming method for those measurements with inevitably noisy interference.
\end{abstract}

\maketitle

\section{I\lowercase{ntroduction}}\label{s:introduction}
%


Compressive sensing  \cite{Candes:06CS, Candes:06CSV2, Candes:08CS, Romberg:08CS,Baraniuk:11Sci}, a new  revolutionary signal processing strategy in information technology, has been recently proposed to extensively relieve the required sampling rate. It should be interesting to apply this idea to extend beamforming techniques  \cite{Beamforming:88} that visualize a signal of interest with a sensor array.
Potential applications can be found in acoustics \cite{ABM:95JASA,ABM:09JASA,Li:08beamforming,Liu:08JASA,Xun:11JASA,Xun:12JASA} that includes those practical test scenarios with poor signal-to-noise ratio (SNR), which could be quite challenging to compressive sensing based beamforming. In an effort to fill this gap, we developed an algorithm of compressive sensing based beamforming, which constitutes the main contribution of this work.


Recently, some compressive sensing based beamforming algorithms have been developed for direction-of-arrival estimation problem \cite{CS:08IEEE, Edelmann:11JASA}. It has been demonstrated in numerical simulations that a large quantity of samples can be saved for most sensors, except the so-called reference sensor that should still satisfy the well-known Nyquist-Shannon sampling rate \cite{CS:08IEEE}.
For practical data from undersea measurements, it was proposed that the sensing matrix should be manipulated to maintain a stable calculation of compressive sensing \cite{Edelmann:11JASA}.

The particular attention of the present article is to develop a working algorithm for compressive sensing in aeroacoustic tests that usually contains strong background noise and broadband interference. The algorithm that we proposed here for compressive sensing beamforming is different to those in the literature. It was tested in this work using either simulated data with various levels of SNR or practical aeroacoustic test data. The results suggest that the proposed algorithm is effective, general and of wide applications.

The rest of the paper is organized as follows. Section~2 introduces the preliminary knowledge of wave model and compressive sensing. Section~3 developed two algorithms of compressive sensing based beamforming. Section~4 examines the proposed algorithms.  A simulation case with a simplified monopole signal is firstly considered. In particular, the effect of measurement noise on compressive sensing is investigated.   Then, the proposed algorithms are applied to a practical aeroacoustic test case to demonstrate general applicability. Finally, Section~5 concludes the present work.

\section{P\lowercase{reliminary knowledge}}\label{s:wavemodel}

\subsection{Wave model}
%

Given a sensory array with $M\,$ microphones, the output $\textbf{y}(t)$ denotes time domain measurements, $\textbf{y} = (y_1 \dots y_i \dots y_M)^T \in  \mathfrak{R}^{M\times1}$ and $(\cdot)^T$ stands for transpose.  For a single signal of interest $s(t) \in \mathfrak{R}^1$ in a free propagation space, using the associated Green's function, we can have
\begin{equation}\label{e:1}
\textbf{y}(t)=\frac{1}{4\pi \textbf{r}}s(t-\tau),  {\tau}=\frac{\textbf{r}}{C},
\end{equation}
\noindent where $C$ is the propagation speed; $\textbf{r} \in \mathfrak{R}^{M\times1}$ are the distances between $s$ and sensors; and  ${\tau}$ is the related sound propagation time delay.

For practical applications, beamforming is generally conducted in the frequency domain.  The frequency domain version of Eq.~(\ref{e:1}) is:
\begin{equation}\label{e:2}
\textbf{Y}({j\omega})=\frac{1}{4\pi \textbf{r}}S({j\omega})e^{-j\omega \tau}=\textbf{G}_v(\textbf{r},j\omega)S({j\omega}),
\end{equation}	
\noindent where $j=\sqrt{-1}$; $\textbf{G}_v \in  \mathfrak{C}^{M\times1}$ is the associated steering vector; $\omega$ is angular frequency; $(j\omega)$ and $(\textbf{r},j\omega)$ are omitted in the following for brevity; $\textbf{Y}$ and $S$ are in the frequency domain. For simplicity, we can write Eq.~(\ref{e:2}) as $\textbf{Y}=\textbf{G}_v S$. The subscript $(\cdot)_v$ suggests  that $\textbf{G}_v$ is a vector.

The situation becomes more complicated for multiple signals of interest plus measurement noise. For clarity, the array output is represented in the scalar form,
\begin{equation}\label{e:scalar}
%
%
\begin{bmatrix}
Y_1\\
\vdots \\
Y_i \\
\vdots\\
Y_M
\end{bmatrix}= \begin{bmatrix}
G_{11} & \dots  & G_{1k} & \dots & G_{1N}\\
 &  & \vdots &  & \\
G_{i1} & \dots & G_{ik} & \dots  & G_{iN} \\
 &  & \vdots  &  & \\
G_{M1} & \dots   &  G_{Mk} & \dots & G_{MN}
\end{bmatrix} \begin{bmatrix}
S_1\\
\vdots\\
S_k \\
\vdots \\
S_N
\end{bmatrix} + \begin{bmatrix}
N_1\\
\vdots\\
N_i\\
\vdots \\
N_M
\end{bmatrix},
\end{equation}
\noindent where $G_{ik}$ is the steering vector between the $i\,$th sensor and the  $k\,$th signal of interest; $Y_i(j\omega)$ is the $i\,$th sensory measurements; $S_k(j\omega)$ is the  $k\,$th signal of interest; and $N_i(j\omega)$ is the collective measurement noise of the $i\,$th sensor.
Potential noise sources include background interference and electronic noise during data acquisition.
For brevity, Eq.~(\ref{e:scalar}) can be written as
\begin{equation}\label{e:SN}
\textbf{Y}=\textbf{G}_{m_I} \textbf{S} + \textbf{N},
\end{equation}	
\noindent where $\textbf{G}_{m_I}  \in \mathfrak{R}^{M \times N}$ is the associated matrix of steering vectors; and $\textbf{S} \in  \mathfrak{C}^{N \times 1}$ and   $\textbf{N}  \in  \mathfrak{C}^{M \times 1} $. Generally, it is assumed that  $\textbf{S}$ and $\textbf{N}$ are of zero-mean and statistically independent. In this work, we define the SNR of the  $k\,$th sensor in  decibels, as the following,
\begin{equation}\label{e:snr}
\mathrm{SNR_{dB}} =  10 \log_{10} \left ( \frac{ | \sum_{i=1}^N G_{ik} S_i  |^2 }{ |N_i|^2 } \right ),
\end{equation}
\noindent where the variables are the same as those in Eq.~(\ref{e:scalar}).

\subsection{Compressive sensing}
%

Candes \textit{et al.} \cite{Candes:06CS} proposed that a perfect reconstruction of a discrete-time signal $\sigma  \in \mathcal{C}^N$ using sub-Shannon sampling rates is possible, as long as $\sigma$  is \textit{sparse} in some Hilbert basis  $\psi \in \mathcal{C}^{N\times N}$, that is, $\sigma= \psi \alpha, \alpha \in \mathcal{C}^N $.  The so-called \textit{sparsity}  means that the number of nonzero entries in $\alpha$  is pretty small, i.e. $|| \alpha ||_0 \ll N$.

According to  compressive sensing theory, we can perform a small number of measurements to collect $\mathbf{y} = \phi \sigma$, where $\mathbf{y} \in  \mathcal{C}^K$ and the sensing matrix $\phi  \in \mathcal{C}^{K\times N}$, in the form of underdetermined linear equations. The sparse signal can then be reconstructed from those $K$ projections by solving an $L_1$ regularization optimization \cite{Candes:06CS},
\begin{equation}\label{e:L1}
\mathrm{arg~min} ~|| \hat{{\alpha}}  ||_{1}, \mathrm {subject~to~} {\phi \psi}\hat{\alpha}=y_i, i=1, ..., K,
\end{equation}
\noindent where $\hat{(\cdot)}$ represents the recovered estimation.


For those measurements polluted by some noise, a closely related programming with an error constraint should be adopted, as the following
\begin{equation}\label{e:L1_V2}
\mathrm{arg~min} ~|| \hat{{\alpha}}  ||_{1}, \mathrm {subject~to~} { || \mathbf{y}-\phi \psi}\hat{\alpha} ||_2 \leq \delta, \delta>0.
\end{equation}
\noindent In this work, $\delta$ is empirically chosen according to the corresponding SNR.

The above programming can be resolved using any available convex optimization tools, such as CVX \cite{Boyd:CVX04}. Once $\hat{\alpha}$ is achieved,  the original signal $\sigma$ can be straightforwardly recovered as $\psi \hat{\alpha}$. The reconstruction error is negligible  with a high probability  if
\begin{equation}\label{e:L1_V4}
K>C_k \cdot M \cdot \mathrm{log} (N/M),
\end{equation}
\noindent where $C_k$ is a universal constant that directly determines the accuracy of the optimization outcomes. A small $C_k$, such as 2, could work if the mutual coherence between $\phi$ and $\psi$ is small. Random projections for $\phi$ is thus recommended in the literature for their general incoherence with respect to most fixed transformation basis $\psi$.

%

%
%


\section{C\lowercase{ompressive sensing beamforming}}\label{s:CSB}



To the best of our knowledge, beamforming work based on compressive sensing is rarely applied to practical applications as yet. Because the only way to validate an algorithm is to apply it for practical applications, the main contribution of this work fills the gap, developing compressive sensing based beamforming algorithms for one of aeroacoustic applications.

%

In this work, signals of interest are presumably regarded as spatially sparse. The same assumption has been adopted in the literature
 \cite{Gurbuz:IEEE12}  for bearing estimation. Then, by checking with Eq.~(\ref{e:L1_V2}) and Eq.~(\ref{e:SN}), we can simply propose a straightforward algorithm of compressive sensing based beamforming,
\begin{equation}\label{e:CS1}
\mathrm{arg~min} ~|| \mathbf{\hat{S}}  ||_{1}, \mathrm {subject~to~} || \mathbf{Y}- \mathbf{G}_{m_I} \mathbf{\hat{S}} ||_2 \leq \delta, \delta>0,
\end{equation}
\noindent where $\delta$ is empirically assigned according to the corresponding SNR, and $\delta=0$ if the measurements are free of noise (i.e., $\mathrm{SNR}=\infty$).  In addition, the beamforming results are generally represented by signal power. Then, the estimated signal power is
\begin{equation}\label{e:CS1V2}
\mathbf{P}_{CSB-I} = || \mathbf{\hat{S}}  ||_{2}.
\end{equation}
For convenience, this compressive sensing beamforming is denoted by CSB-I in the following.


In this work, we developed a new compressive sensing beamforming algorithm based on so-called cross spectrum matrix (also known as covariance matrix or cross-spectral density matrix).  The definition is
\begin{equation}\label{e:cov}
\textbf{R} = \mathrm{E}\left \{ \textbf{Y}\textbf{Y}^*\right \},
\end{equation}	
\noindent  which can be approximated by
\begin{equation}
\label{e:scov1} \hat{\mathbf{R}}  \approx  \frac{1}{K}{}\sum_{k=1}^{K} \textbf{Y}\textbf{Y}^*,
\end{equation}
\noindent  where $K$ is the number of sampling blocks. For statistical confidence,
the associated sampling duration should be much larger than the the period of signal of interest.

The associated algorithm is called CSB-II throughout this article. Its derivation is as the following. From Eq.~(\ref{e:scalar}), we have
\begin{equation}\label{e:CSBII}
\mathbf{{R}}_V
= \begin{bmatrix}
G_{11} G_{11}^*  & G_{12} G_{12}^*     & \dots & G_{1N} G_{1N}^*\\
G_{11} G_{21}^*  & G_{12} G_{22}^*     & \dots & G_{1N} G_{2N}^*\\
 &  & \vdots    & \\
G_{M1} G_{M1}^*  & G_{M2} G_{M2}^*     & \dots & G_{MN} G_{MN}^*\\
\end{bmatrix}
\mathbf{P} +
\mathbf{Q} \newline
= \mathbf{G}_{m_{II}} \mathbf{P} + \mathbf{Q},
%
\end{equation}
\noindent where  $\mathbf{{R}}_V = \left(\mathrm{E}[Y_1Y_1^*], \mathrm{E}[Y_1 Y_2^*], \cdots, \mathrm{E}[Y_M Y_M^*] \right)^T  \in  \mathfrak{C}^{M^2 \times N} $  is a vector that can be approximately achieved by reshaping $\mathbf{\hat{R}}$ [Eq.~(\ref{e:scov1})]; the symbol of $(\cdot)^*$ denotes the conjugate transpose; $\mathbf{P}= [S_1 S_1^*, S_2 S_2^*, \cdots, S_N S_N^*]^T \in  \mathfrak{R}^{N\times1}$, where $S_i$ and $S_k$ ($i \ne k$) are incoherent; and $\mathbf{Q}$ is the vertical vector form of $\mathrm{E}[\mathbf{N} \mathbf{N}^*] \in  \mathfrak{C}^{M ^2\times1}$.   Then, the proposed CSB-II algorithm works by solving
\begin{equation}\label{e:CS2}
\mathrm{arg~min} ~|| \hat{\mathbf{P}}  ||_{1}, \mathrm {subject~to~} \hat{\mathbf{P}} >0,  || \mathbf{\hat{R}}_V- \mathbf{G}_{m_{II}} \hat{\mathbf{P}} ||_2 \leq \delta, \delta>0.
\end{equation}
\noindent As a result, the estimated signal power is
 \begin{equation}\label{e:CSII}
\mathbf{P}_{CSB-II} = \mathbf{\hat{P}}.
\end{equation}
%


 In summary, both the algorithms developed in this work are started by,

Step 0: Collect measurements and obtain $\mathbf{Y}$ by performing Fourier transform.

Then, the CSB-I algorithm is conducted as the following,

Step 1: Prepare $\mathbf{G}_{m_{I}}$ using Eqs.~(\ref{e:2})--(\ref{e:SN}).

Step 2: Calculate the CSB-I beamforming with Eqs.~(\ref{e:CS1})--(\ref{e:CS1V2}). Done.

On the other hand, the steps of the CSB-II algorithm  include

Step 1: Prepare $\mathbf{G}_{m_{II}}$ using Eq.~(\ref{e:CSBII}).

Step 2: Prepare $\mathbf{R}_V$ by reshaping $\hat{\mathbf{R}}$ that is calculated with Eq.~(\ref{e:scov1}).

Step 3: Perform $l_1$-minimization  and achieve the CSB-II beamforming results, using Eqs.~(\ref{e:CS2})--(\ref{e:CSII}). Done.

\section{R\lowercase{results and discussion}}\label{s:results}
Test data achieved in our previous experiments \cite{Huang:10JSV} is used to examine the proposed algorithms (CSB-I and CSB-II). The experimental setup is briefly introduced for the completeness of this paper. The experiments were conducted in an anechoic chamber facility ($9.15\,$m $\times$ $9.15\,$m $\times$ $7.32\,$m) at ISVR, University of Southampton. Figure 1 shows the setup. A nozzle ($500\,$mm $\times$ $350\,$mm) connecting to a plenum chamber produces a jet flow of $U_\infty= 30\,$m/s. An airframe model representing a part of a landing gear was studied in experiments. The corresponding Reynolds number based on the cylinder diameter is $2.1 \times 10^5$.

\begin{figure}[h!]
\begin{center}
    \includegraphics[width=150mm]{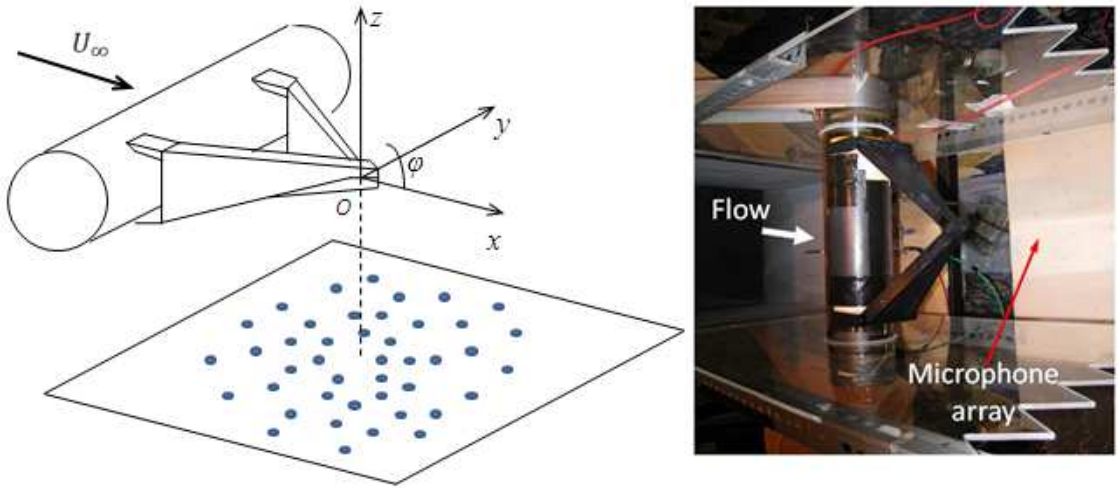}
    \caption { The setup of experimental units.    } \label{f:spectrum}
\end{center}
\end{figure}
An array of 56 electret microphones (Panasonic WM-60A) is placed on the ground, underneath the test model. The coordinates shown in Fig. 1 are used throughout the rest of the paper. The distance between the array and the model is $0.7\,$m. The center of the array is aligned with the origin.
The sensitivity of each microphone is  $-45 \pm 5\,$dBV/Pa. The frequency response of each WM-60A microphone is calibrated to decrease amplitude and phase deviations from a B\&K 4189 microphone. The layout of the microphones is a multi-arm spiral line, which is \textit{de facto} adopted in acoustic tests (pp.118--128, in the reference \cite{ArrayBook1}).

Before working on those experimental data, we first conducted a simulation case to quantity the performance of the proposed algorithms. In the simulation, we assume
free space propagation for a monopole tonal sources that locates at the origin, $1\,$m away from the array. The frequency of the source is $5\,$kHz. For this simple  case, we just use 10 microphones, which are randomly chosen in the array, to yield the beamforming results.  In addition to this sound wave, we assume that each sensor also perceive a white noise, which could come from background noise or electronic noise.
Various SNR levels, from $\infty$ to $-10\,$dB, have been tested. According to Eq.~(\ref{e:snr}), $\mathrm{SNR}=\infty$ suggests a negligible noise; $\mathrm{SNR}=0\,$dB suggests that the power from the monopole signal equals the power from the background noise; and $\mathrm{SNR}=-10\,$dB suggests that the power of the background noise is ten times greater than the power of the monopole signal.

\begin{figure}[h!]
\begin{center}
    \includegraphics[width=150mm]{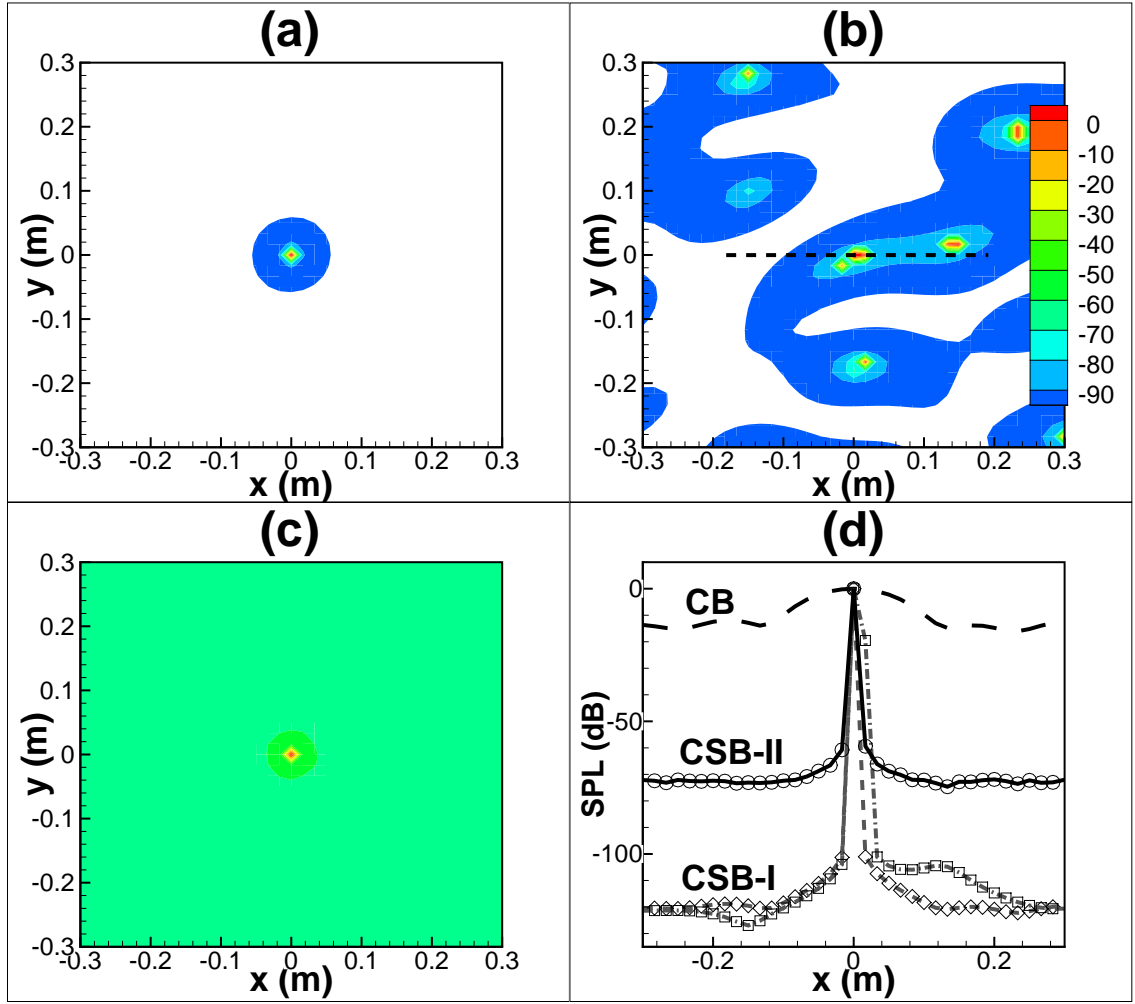}
    \caption { Beamforming results of the monopole case at $5\,$kHz, where: (a) $\mathrm{SNR}=\infty$, CSB-I; (b) $\mathrm{SNR}=-10\,$dB, CSB-I; (c) $\mathrm{SNR}=-10\,$dB, CSB-II; and (d) the $x$-axial only performance, ($--$) the CB result for $\mathrm{SNR}=-10\,$dB, ($\diamond$) the CSB-I result for
    $\mathrm{SNR}=\infty$,  ($\square$)  the CSB-I result for $\mathrm{SNR}=-10\,$dB, and ($\circ$) the CSB-II result for $\mathrm{SNR}=-10\,$dB.
    The contour levels in (a-c) are set between $-100\,$dB and $0\,$dB. The levels below $-100\,$dB are cut off.
     } \label{f:spectrum}
\end{center}
\end{figure}

Figures~\ref{f:spectrum}(a-b) show the normalized beamforming results using the CSB-I algorithm.
The dynamic range of the CSB-I results is more than $100\,$dB and those data smaller than $-100\,$dB is cut off for clarity of the figures.
In Fig.~\ref{f:spectrum}(a), the simulated measurements are free of background noise (i.e., $\mathrm{SNR}=\infty$). It can be seen that the CSB-I algorithm perfectly capture the desired signal with very fine resolution and nice sidelobe rejection. However, as the value of SNR decreases, we found that the CSB-I algorithm fails to output reasonable results. For example, when $\mathrm{SNR}=-10\,$dB, Fig.~\ref{f:spectrum}(b) shows that false signal sources scatter on the entire imaging domain. In contrast,  Fig.~\ref{f:spectrum}(c) shows that the CSB-II algorithm is still able to capture the mainlobe as well as  maintain a good sidelobe rejection. The dynamic range is however diminished to $60\,$dB that is still quite satisfactory.

\begin{figure}[h!]
\begin{center}
    \includegraphics[width=150mm]{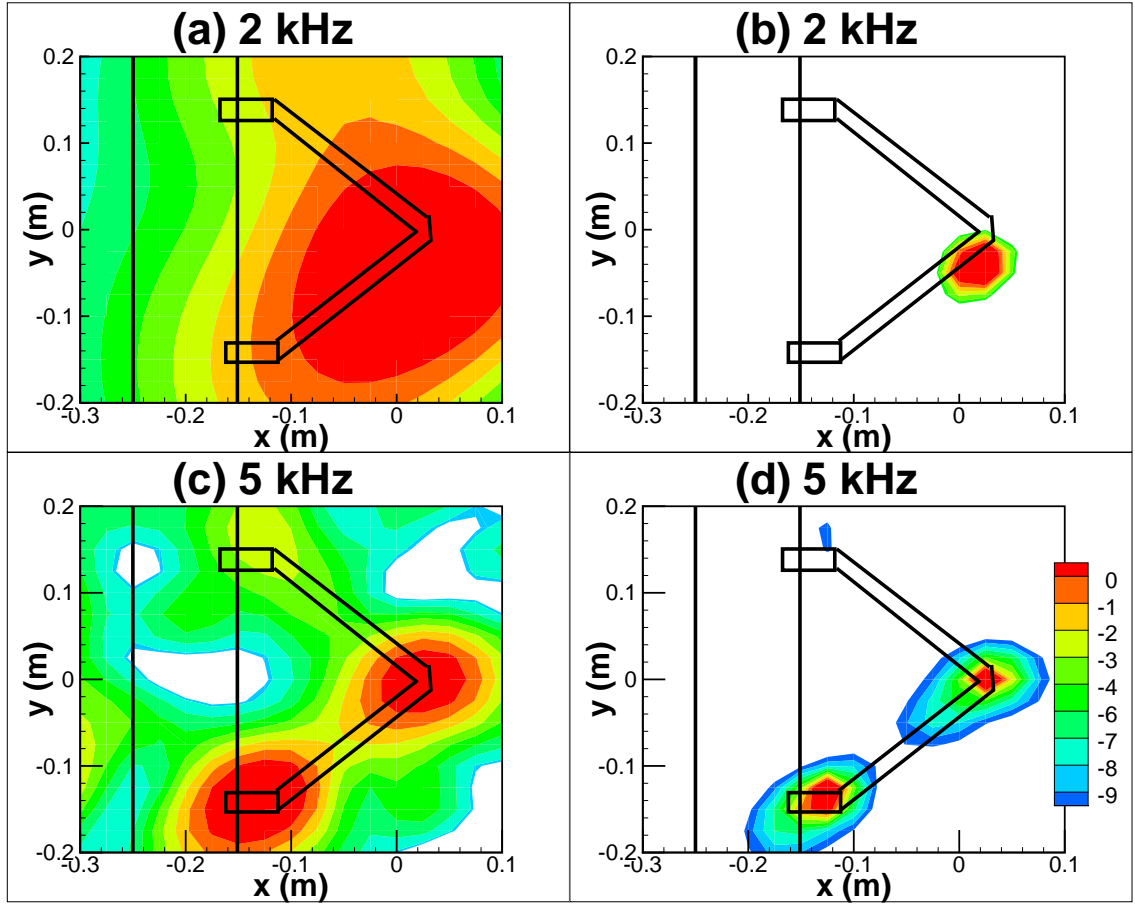}
    \caption { Acoustic images, (a-b) $2\,$kHz, (c-d) $5\,$kHz, (a)(c) using the CB algorithm, and (b)(d) using the CSB-II algorithm.} \label{f:torque}
\end{center}
\end{figure}
Figure~\ref{f:spectrum}(d) examines the $x$-axial performance on the dashed line, which has been shown in Fig.~\ref{f:spectrum}(b).  In comparing CSB-I and CSB-II algorithms with conventional beamforming (CB), the following expression is used,
\begin{equation}\label{e:beamforming}
{\sigma}_{CB}=\mathbf{W}^*\mathbf{\hat{R}} \mathbf{W}, \mathbf{W}=(\mathbf{G}_v ^* \mathbf{G}_v)^{-1} \mathbf{G}_v.
\end{equation}
\noindent The above CB algorithm is narrowband and only for a single gridpoint. We have to scan each gridpoint of the imaging plane using this algorithm to yield the desired images at frequency ranges of interest. It can be seen that the CB algorithm produces a very broad mainlobe. The associated dynamic range is slightly over $10\,$dB. Then, Fig.~\ref{f:spectrum}(d) clearly identify the distinctive performance of CSB algorithms, in terms of the resolution and dynamic range. In addition, the CSB-II algorithm can suppress detrimental  interference of noisy measurements to some extent. In this very simple case with an idealized monopole source, the CSB-II algorithm fails to produce correct results if $\mathrm{SNR}$ is smaller than $-15\,$dB. On the other hand, the CSB-I algorithm fails quickly when $\mathrm{SNR}$ is just $0\,$dB. As a result, only the CSB-II algorithm is applied in the following practical test.


Most compressive sensing works are validated using simulation results. Very few beamforming results from compressive sensing  can be found in the literature for practical experimental data. It should be interesting to
apply the proposed method to practical aeroacoustic test data, which consists of multiple broadband noise sources. In this work, a bluff body model that represents the main part of a landing gear is used. The CSB-II results are compared to those obtained with the CB algorithm at various frequencies.
Figure~\ref{f:torque} shows some results at $2\,$kHz and $5\,$kHz. The contour levels are between $-10\,$dB and $0\,$dB.		
Compared to the CB results, the CSB-II results have a better resolution (with narrow mainlobes) and smaller sidelobe levels. In short, the imaging quality is improved with the proposed compressive sensing based beamforming method.

The CSB-II algorithm was developed in MATLAB and computed on a laptop with an Intel i5 CPU (@$1.7\,$GHz) and $4\,$GB $1333\,$MHz DDR3 memory. The CSB-II algorithm spends $450\,$s for the case including 56 sensors and 1600 imaging gridpoints. In addition, the calculation cost of the CSB-I algorithm is much less and negligible. It is worthwhile to mention that the code is not extensively optimized.

%

\section{S\lowercase{ummary}}\label{s:summary}
Compressive sensing is the newly emerging method in information technology that could significantly impact acoustic research and applications. In this article, we firstly introduced the fundamentals of compressive sensing theory.  After that, we implemented two different compressive sensing based beamforming algorithms (CSB-I and CSB-II). Both algorithms are proposed for those presumably spatially sparse and incoherent signals.

The two algorithms are examined using a simple simulation case and a practical aeroacoustic test case. The simulation case clearly shows that  the CSB-I algorithm is quite sensitive to the sensing noise. The CSB-II algorithm, on the other hand, is more robust to noisy measurements. The results by CSB-II at $\mathrm{SNR}=-10\,$dB are reasonable with good resolution and sidelobe rejection. Although the inherent reason is not discussed in this work, we believe it has connection with the so-called restricted isometry property \cite{Candes:06CS}. Detailed analysis id beyond the scope of this paper.

The proposed method was then successfully evaluated and demonstrated in the numerical simulations. The sound source considered in the simulation case is an idealised monopole. Few results for practical experimental data can be found in the literature. This work develops compressive sensing based beamforming algorithms specifically for aeroacoustic tests and applies it to the practical data in an effort to fill this gap. The CSB-II algorithm is applied to experimental data acquired in an anechoic chamber facility. The results suggest that the proposed CSB-II algorithm is robust to potential interference in practical tests, and can produce an acoustic image with a significant improvement of resolution. As a result, the classical deconvolution post-processing can be omitted, or as suggested in the reference \cite{Xun:12JASA}, the post-processing time can be extensively improved.



\section*{Acknowledgement}
This research was supported by the NSF Grant of China (Grants No. 11172007).  The experiments were conducted at ISVR, University of Southampton. We acknowledge Professor Xin Zhang for his support in the experiments.

\bibliographystyle{elsarticle-num-names}
\bibliography{references}%

\begin{thebibliography}{18}
\providecommand{\natexlab}[1]{#1}
\providecommand{\url}[1]{\texttt{#1}}
\providecommand{\urlprefix}{URL }
\expandafter\ifx\csname urlstyle\endcsname\relax
  \providecommand{\doi}[1]{doi:\discretionary{}{}{}#1}\else
  \providecommand{\doi}[1]{doi:\discretionary{}{}{}\begingroup
  \urlstyle{rm}\url{#1}\endgroup}\fi
\providecommand{\bibinfo}[2]{#2}

\bibitem[{Candes et~al.(2006)Candes, Romberg, and Tao}]{Candes:06CS}
\bibinfo{author}{E.~J. Candes}, \bibinfo{author}{J.~Romberg},
  \bibinfo{author}{T.~Tao}, \bibinfo{title}{Robust Uncertainty Principles:
  Exact Signal Reconstruction From Highly Incomplete Frequency Information},
  \bibinfo{journal}{IEEE Transaction on Information Theory}
  \bibinfo{volume}{52}~(\bibinfo{number}{2}) (\bibinfo{year}{2006})
  \bibinfo{pages}{489--509}.

\bibitem[{Candes(2006)}]{Candes:06CSV2}
\bibinfo{author}{E.~J. Candes}, \bibinfo{title}{Near-Optimal Signal Recovery
  From Random Projections: Universal Encoding Strategies?},
  \bibinfo{journal}{IEEE Transaction on Information Theory}
  \bibinfo{volume}{52}~(\bibinfo{number}{12}) (\bibinfo{year}{2006})
  \bibinfo{pages}{5406--5425}.

\bibitem[{Candes and Wakin(2008)}]{Candes:08CS}
\bibinfo{author}{E.~J. Candes}, \bibinfo{author}{M.~B. Wakin},
  \bibinfo{title}{An Introduction To Compressive Sampling},
  \bibinfo{journal}{IEEE Signal Processing Magzaine}
  \bibinfo{volume}{25}~(\bibinfo{number}{2}) (\bibinfo{year}{2008})
  \bibinfo{pages}{21--30}.

\bibitem[{Romberg(2008)}]{Romberg:08CS}
\bibinfo{author}{J.~Romberg}, \bibinfo{title}{Imaging via Compressive
  Sampling}, \bibinfo{journal}{IEEE Signal Processing Magzaine}
  \bibinfo{volume}{25}~(\bibinfo{number}{2}) (\bibinfo{year}{2008})
  \bibinfo{pages}{14--20}.

\bibitem[{Baraniuk(2011)}]{Baraniuk:11Sci}
\bibinfo{author}{R.~G. Baraniuk}, \bibinfo{title}{More Is Less: Signal
  Processing and the Data Deluge}, \bibinfo{journal}{Science}
  \bibinfo{volume}{331}~(\bibinfo{number}{11}) (\bibinfo{year}{2011})
  \bibinfo{pages}{717--719}.

\bibitem[{{Van~Veen} and Buckley(1988)}]{Beamforming:88}
\bibinfo{author}{B.~D. {Van~Veen}}, \bibinfo{author}{K.~M. Buckley},
  \bibinfo{title}{Beamforming: A Versatile Approach to Spatial Filtering},
  \bibinfo{journal}{IEEE ASSP Magazine}
  \bibinfo{volume}{5}~(\bibinfo{number}{2}) (\bibinfo{year}{1988})
  \bibinfo{pages}{4--24}.

\bibitem[{Gramann and Mocio(1995)}]{ABM:95JASA}
\bibinfo{author}{R.~A. Gramann}, \bibinfo{author}{J.~W. Mocio},
  \bibinfo{title}{Aeroacoustic measurements in wind tunnels using adaptive
  beamforming methods}, \bibinfo{journal}{Journal of Acoustical Society of
  America} \bibinfo{volume}{97}~(\bibinfo{number}{6}) (\bibinfo{year}{1995})
  \bibinfo{pages}{3694--3701}.

\bibitem[{Cho and Roan(2009)}]{ABM:09JASA}
\bibinfo{author}{Y.~T. Cho}, \bibinfo{author}{M.~J. Roan},
  \bibinfo{title}{Adaptive near-field beamforming techniques for sound source
  imaging}, \bibinfo{journal}{Journal of Acoustical Society of America}
  \bibinfo{volume}{125}~(\bibinfo{number}{2}) (\bibinfo{year}{2009})
  \bibinfo{pages}{944--957}.

\bibitem[{Yardibi et~al.(2008)Yardibi, Li, Stoica, and
  Cattafesta}]{Li:08beamforming}
\bibinfo{author}{T.~Yardibi}, \bibinfo{author}{J.~Li},
  \bibinfo{author}{P.~Stoica}, \bibinfo{author}{L.~N. Cattafesta},
  \bibinfo{title}{Sparsity constrained deconvolution approaches for acoustic
  source mapping}, \bibinfo{journal}{Journal of Acoustical Society of America}
  \bibinfo{volume}{123}~(\bibinfo{number}{5}) (\bibinfo{year}{2008})
  \bibinfo{pages}{2631--2642}.

\bibitem[{Liu et~al.(2008)Liu, Quayle, Dowling, and Sijtsma}]{Liu:08JASA}
\bibinfo{author}{Y.~Liu}, \bibinfo{author}{A.~R. Quayle},
  \bibinfo{author}{A.~P. Dowling}, \bibinfo{author}{P.~Sijtsma},
  \bibinfo{title}{Beamforming correction for dipole measurement using
  two-dimensional microphone arrays}, \bibinfo{journal}{Journal of Acoustical
  Society of America} \bibinfo{volume}{124}~(\bibinfo{number}{1})
  (\bibinfo{year}{2008}) \bibinfo{pages}{182--191}.

\bibitem[{Bai and Huang(2011)}]{Xun:11JASA}
\bibinfo{author}{L.~Bai}, \bibinfo{author}{X.~Huang},
  \bibinfo{title}{Observer-based Beamforming Algorithm for Acoustic Array
  Signal Processing}, \bibinfo{journal}{Journal of Acoustical Society of
  America} \bibinfo{volume}{130}~(\bibinfo{number}{6}) (\bibinfo{year}{2011})
  \bibinfo{pages}{3803--3811}.

\bibitem[{Huang et~al.(2012)Huang, Bai, Vinogradov, and Peers}]{Xun:12JASA}
\bibinfo{author}{X.~Huang}, \bibinfo{author}{L.~Bai},
  \bibinfo{author}{I.~Vinogradov}, \bibinfo{author}{E.~Peers},
  \bibinfo{title}{Adaptive Beamforming for Array Signal Processing in
  Aeroacoustic Measurements}, \bibinfo{journal}{Journal of Acoustical Society
  of America} \bibinfo{volume}{131}~(\bibinfo{number}{3})
  (\bibinfo{year}{2012}) \bibinfo{pages}{2152--2161}.

\bibitem[{Gurbuz et~al.(2008)Gurbuz, McClellan, and Cevher}]{CS:08IEEE}
\bibinfo{author}{A.~C. Gurbuz}, \bibinfo{author}{J.~H. McClellan},
  \bibinfo{author}{V.~Cevher}, \bibinfo{title}{A Compressive Beamforming
  Method}, \bibinfo{note}{iEEE Acoustics, Speech and Signal Processing},
  \bibinfo{year}{2008}.

\bibitem[{Edelmann and Gaumond(2011)}]{Edelmann:11JASA}
\bibinfo{author}{G.~F. Edelmann}, \bibinfo{author}{C.~F. Gaumond},
  \bibinfo{title}{Beamforming using compressive sensing},
  \bibinfo{journal}{Journal of Acoustical Society of America}
  \bibinfo{volume}{130}~(\bibinfo{number}{4}) (\bibinfo{year}{2011})
  \bibinfo{pages}{EL232--EL237}.

\bibitem[{Boyd and Vandenberghe(2004)}]{Boyd:CVX04}
\bibinfo{author}{S.~Boyd}, \bibinfo{author}{L.~Vandenberghe},
  \bibinfo{title}{Convex Optimization}, \bibinfo{publisher}{Cambridge
  University Press New York}, \bibinfo{year}{2004}.

\bibitem[{Gurbuz et~al.(2012)Gurbuz, Cevher, and McClellan}]{Gurbuz:IEEE12}
\bibinfo{author}{A.~C. Gurbuz}, \bibinfo{author}{V.~Cevher},
  \bibinfo{author}{J.~H. McClellan}, \bibinfo{title}{Bearing Estimation via
  Spatial Sparsity using Compressive Sensing}, \bibinfo{journal}{IEEE
  Transactions on Aerospace and Electronic Systems}
  \bibinfo{volume}{48}~(\bibinfo{number}{2}) (\bibinfo{year}{2012})
  \bibinfo{pages}{1358--1369}.

\bibitem[{Huang et~al.(2010)Huang, Zhang, and Li}]{Huang:10JSV}
\bibinfo{author}{X.~Huang}, \bibinfo{author}{X.~Zhang},
  \bibinfo{author}{Y.~Li}, \bibinfo{title}{Broadband Flow-Induced Sound Control
  using Plasma Actuators}, \bibinfo{journal}{Journal of Sound and Vibration}
  \bibinfo{volume}{329}~(\bibinfo{number}{13}) (\bibinfo{year}{2010})
  \bibinfo{pages}{2477--2489}.

\bibitem[{Mueller(2002)}]{ArrayBook1}
\bibinfo{author}{T.~J.~E. Mueller}, \bibinfo{title}{Aeroacoustic Measurements},
  \bibinfo{publisher}{Springer, Germany}, \bibinfo{year}{2002}.

\end{thebibliography}

\end{document}